\documentclass[letterpaper
]{article} 
\usepackage[]{aaai25}  
\usepackage{times}  
\usepackage{helvet}  
\usepackage{courier}  
\usepackage[hyphens]{url}  
\usepackage{graphicx} 
\usepackage{natbib}  
\usepackage{caption} 
\usepackage{url}            
\usepackage{booktabs}       
\usepackage{amsfonts}       
\usepackage{nicefrac}       
\usepackage{microtype}      
\usepackage{amsthm}
\usepackage{caption}
\usepackage{subcaption}
\usepackage{graphics} 
\usepackage{epsfig} 
\usepackage{times} 
\usepackage{amsmath} 
\usepackage{amssymb}  
\usepackage{makeidx}
\usepackage{float}
\usepackage{booktabs}

\usepackage{mathrsfs}
\usepackage{enumerate}
\usepackage{multicol}

\usepackage{algorithm}
\usepackage{algpseudocode}
\usepackage{braket}
\usepackage{easyReview}
\usepackage{mathrsfs}  
\usepackage{xr} 
\usepackage{multicol}
\usepackage{float}
\usepackage[nolist]{acronym}
\usepackage{xcolor}

\usepackage{lipsum} 

\usepackage{cleveref} 

\usepackage[acronym]{glossaries} 

\usepackage[compat=0.4]{yquant}
\usetikzlibrary{quotes}

\usepackage{scalerel}
\usepackage{newfloat}
\usepackage{listings}
\DeclareCaptionStyle{ruled}{labelfont=normalfont,labelsep=colon,strut=off} 
\floatstyle{ruled}
\newfloat{listing}{tb}{lst}{}
\floatname{listing}{Listing}
\title{Predicting Chaotic Systems with Quantum Echo-state Networks}

\author { 
   Erik Connerty\textsuperscript{\rm 1},
   Ethan N. Evans\textsuperscript{\rm 2},
   Gerasimos Angelatos\textsuperscript{\rm 3},
   Vignesh Narayanan\textsuperscript{\rm 1}
}
\affiliations {
   \textsuperscript{\rm 1}College of Engineering and Computing, University of South Carolina - Columbia\\
   \textsuperscript{\rm 2}Naval Surface Warfare Center, Panama City Division\\
   \textsuperscript{\rm 3}RTX BBN Technologies, Cambridge, MA\\
   erikc@cec.sc.edu,
   ethan.n.evans.civ@us.navy.mil,
   gerasimos.angelatos@rtx.com,
    vignar@sc.edu
   }
\begin{document}

\maketitle

\begin{abstract}
    Recent advancements in artificial neural networks have enabled impressive tasks on classical computers, but they demand significant computational resources. 
    While quantum computing offers potential beyond classical systems, the advantages of quantum neural networks (QNNs) remain largely unexplored. In this work, we present and examine a quantum circuit (QC) that implements and aims to improve upon the classical echo-state network (ESN), a type of reservoir-based recurrent neural networks (RNNs), using quantum computers.  Typically, ESNs consist of an extremely large reservoir that learns high-dimensional embeddings, enabling prediction of complex system trajectories. Quantum echo-state networks (QESNs) aim to reduce this need for prohibitively large reservoirs by leveraging the unique capabilities of quantum computers, potentially allowing for more efficient and higher performing time-series prediction algorithms. 
    The proposed QESN can be implemented on any digital quantum computer implementing a universal gate set, and does not require any sort of stopping or re-initialization of the circuit, allowing continuous evolution of the quantum state over long time horizons. 
    We conducted simulated QC experiments on the chaotic Lorenz system, both with noisy and noiseless models, to demonstrate the circuit's performance and its potential for execution on noisy intermediate-scale quantum (NISQ) computers.
    \end{abstract}

\section{Introduction}\label{sec: Introduction}
    Quantum computers are poised to offer a unique framework for the adaptation of neural networks and all of their offshoots into quantum versions of their classical representations. Due to the inherent difficulties of implementing a large number of quantum gates and the short coherence times of current noisy-intermediate scale quantum (NISQ) hardware \cite{gill2022quantum}, only limited amounts of research have been conducted in the field of gate-based quantum reservoir computers (QRCs). Our work aims to advance some of these challenges by presenting a scalable recurrent neural network (RNN) architecture for both present NISQ and future fault-tolerant quantum computers by leveraging design-principles from the classical echo-state networks (ESN).
    
   ESNs \cite{jaeger2001}, a class of RNNS, offer a unique way of handling time-series data. Specifically, without the need for backpropagation techniques or vast amounts of labeled data, ESNs can perform seemingly impossible tasks with very little hyperparameter tuning. Moreover, sparsely connected hidden layers within ESNs \cite{miao2022interpretable} are a desirable feature that can reduce the gate complexity of the quantum circuit (QC) and may prove to be a promising architecture to implement on the NISQ computers available in the present day. However, developing a quantum version of a classical machine-learning algorithm very rarely is accomplished by converting instructions over one-to-one \cite{evans2024learningsasquatchnovelvariational}. 
    For example, in the context of ESNs, the creation of a \emph{fading memory} (called the``echo-state property'') and \emph{nonlinearity} in quantum ESNs (QESNs) are important challenges to overcome due to the inherent difficulty of quantum coherence times \cite{gill2022quantum} and state measurement collapse, as well as the inherent unitary evolution of closed quantum systems seemingly preventing the implementation of nonlinear operations. 

\textbf{Related Works:}
Previous research has examined the creation of functional quantum RNNs (QRNNs) as well as QRCs. One such approach, detailed in \citet{LI2023148}, developed a QRNN trained through the backpropogation algorithm, which demands large number of circuit evaluations, and may suffer from vanishing gradients in a similar fashion to classical RNNs. In addition, works such as \cite{kornjača2024largescalequantumreservoirlearning,Govia_2022,zhu2024practicalscalablequantumreservoir, dudas2023, Mujal2023} examined using analog quantum computers for time-series prediction in an open quantum system, but these methodologies are restricted to hardware-specific, specialized non-universal devices. Most recently, \citet{Hu2024} developed a methodology for running a QC seemingly indefinitely without intermediate halts or resets. Deterministic resets of a subset of qubits prevent information scrambling and thermalization, and enable time-invariant fading memory {\cite{Hu2024}}. Motivated by this ``measure and reset'' paradigm, in this paper, we develop a QESN circuit, which incorporates sparsely connected randomly initialized reservoir layer with a more robust and tunable entangling scheme for predicting time-series data. 

\textbf{Proposed Method:}
We propose a near term NISQ ready algorithm that can leverage the large Hilbert space \cite{PhysRevResearch.4.033007} afforded by a set of qubits to produce a rich feature space that is suitable for predicting complex dynamics. Our QESN algorithm features a fully connected input layer utilizing a ``context window'' for data input, a novel entanglement scheme using ``data reuploading'' \cite{P_rez_Salinas_2020}, and completely randomly generated and sparse weights throughout both the hidden layer and input layer.  Figure \ref{fig:QRC_Pipeline} details the entire proposed pipeline, 
showing how classical data is input into the QESN, evolved through a quantum channel, and then sampled to generate features for a classical optimization problem.

In the proposed QESN, qubits are grouped into \emph{memory} and \emph{readout} registers. 
{We implement a method for embedding a context window of data using a fully connected input layer. This embedding technique maps arbitrary sized inputs into three Euler angles for each qubit, and allows us to decouple the size of the reservoir from the dimension of the input}. Single qubit rotations interspersed with two-qubit entangling gates are included to introduce nonlinear mappings from input to output, as well as nonlinear combinations of previous inputs with current inputs (i.e., memory) \cite{chu2022qmlperrortolerantnonlinearquantum, Govia_2022} and allow the necessary transfer of information from the memory register to the readout register. By selectively measuring only the readout qubits and then deterministically resetting them to an unbiased state, we are able to run the circuit indefinitely while retaining a fading memory \cite{Hu2024}, an important feature of classical reservoir networks. The sparsely allocated weights on the parameterized entangling gates decrease the number of gate errors, as gates that are zeroed out are removed from the circuit altogether. At each timestep, a probability distribution is sampled from the reservoir to create features for a classical least-squares regression.

\section{Preliminaries and Notation}\label{sec:notations}
The input data is denoted by $X = \{x_1, \dots, x_t, \dots, x_N\}$, with $x_t \in \mathbb{R}^d$ and $t \in T$, where $T = \{1, \dots, N\}$. We encode the classical data onto $n_q$ total qubits, which must be even, using a context window $X_c = \{x_{t-c}, \dots, x_t\}$ of length $c$. We utilize a re-uploading scheme to introduce nonlinearity with $r$ repeated circuit blocks, and treat $r$ as a tunable hyperparameter that can increase the amount of nonlinearity in the circuit. 
The $\textit{Memory}$ and $\textit{Readout}$ registers consist of $\frac{n_q}{2}$ qubits each. The set of pairs of readout qubits are denoted as $P$ with $P_i$ denoting the $i$th element in the pair, and the set of pairs of just memory qubits are denoted as $M$ with $M_i$ similarly denoting the $i$th element in the pair of memory qubits. We also introduce sparsity parameter $\kappa \in [0, 1]$. 

\begin{figure*}
    \centering
    \includegraphics[width=\linewidth]{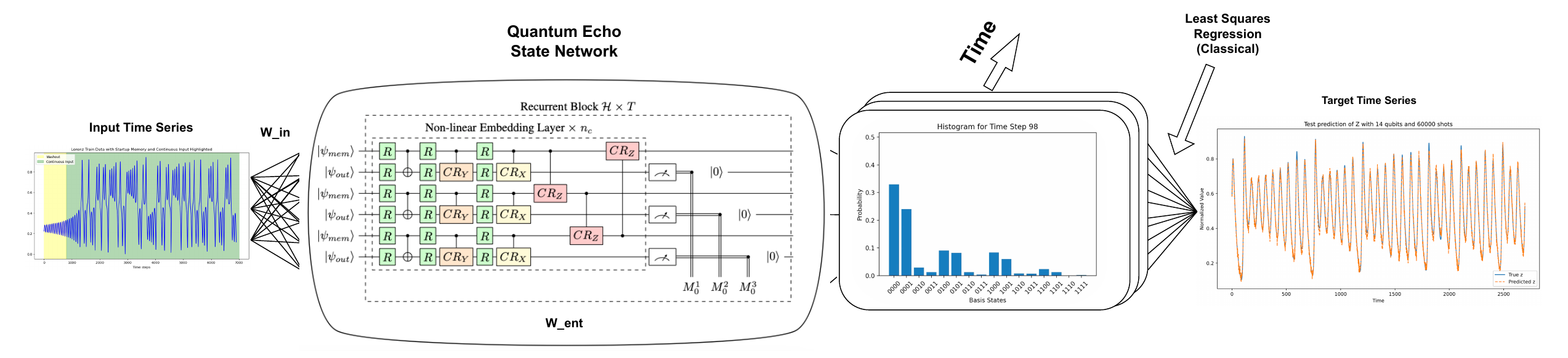}{}
    \caption{\footnotesize Proposed QESN framework. Data is input as a sliding context window over time through a fully connected classical layer, and then passed through a dynamic quantum reservoir that is repeatedly sampled to generate a probability distribution over the $2^\frac{n_q}{2}$ Hilbert space at each time step. With the collected probability distributions, a classical least-squares optimization process fits the reservoir states to a target signal.}
    \label{fig:QRC_Pipeline}
\end{figure*}

Random weight matrices $\mathbf{W}_\text{in} \in \mathbb{R}^{cd \times n_q  \times 3}$
,$\mathbf{W}_\text{bias} \in \mathbb{R}^{n_q \times 1}$, and $\mathbf{W}_\text{ent} \in \mathbb{R}^{n_q/2 \times 2}$, and are used for rotations and their biases, as well as entanglement gates between qubits, respectively. The rotation angles $\alpha$, $\beta$, and $\gamma$ are computed from the context window and the weight matrices $\mathbf{W}_\text{in}$ and $\mathbf{W}_\text{bias}$. Rotation gates are arbitrary single qubit rotations applied to a qubit $\mathbf{q}$ with Euler angles represented as $\mathbf{R(q, \alpha, \beta, \gamma)}$. The gates used in the QC (shown as part of Figure \ref{fig:QRC_Pipeline}) include Controlled-X (C-NOT), Controlled-RY (CRY), Controlled-RX (CRX), and Controlled-RZ (CRZ).

\subsection{System Architecture}\label{subsec:system-architecture}

Here we give the pseudocode for the generation of the QESN circuit in Algorithm \ref{qesnalgo}.
\begin{algorithm}
\caption{QESN Algorithm}
\small
\begin{algorithmic}
\Procedure{Create\_Quantum\_Circuit}{$\mathbf{X}, n_q, c, r, \kappa$}
    \State \textbf{Initialize:} 
    \State Create \textit{Memory} and \textit{Readout} registers with $\frac{n_q}{2}$.
    \State Initialize all qubits to $\ket{0}$
    \State Initialize random weights $\mathbf{W}_\text{in}$ and $\mathbf{W}_\text{ent}$.
    \State Apply sparsity to $\mathbf{W}_\text{ent}$ using $\kappa$
    \For{$t = c$ \textbf{to} $N-c$}
        \State Extract context window $\mathbf{X_c} \gets \mathbf{X}[t-c:t]$
        \For{$j = 1$ \textbf{to} $r$}
            \State \textbf{Parallel operations on qubit pairs}:
            \State $\alpha_i, \beta_i, \gamma_i \gets W_{in}^{P_i} \cdot \mathbf{X_c} + W_{bias}^{P_i},\quad$ $i=1,2$
            \State $\epsilon_1,\epsilon_2 \gets W_{ent}^P$
            \State Apply rotation: $R({P_i}, \alpha_i, \beta_i, \gamma_i)$, $\quad i=1,2$  
            \State Apply C-NOT gate: C-NOT$(P_1, P_2)$
            \State Apply rotation: $R(P_i, \alpha_i, \beta_i, \gamma_i)$, $\quad i=1,2$ 
            \State Apply controlled-RY gate: CRY$(P_1, P_2, \epsilon_1)$
            \State Apply rotation: $R(P_i, \alpha_i, \beta_i, \gamma_i)$, $\quad i=1,2$ 
            \State Apply controlled-RX gate: CRX$(P_1, P_2, \epsilon_2)$
            \State \textbf{Sequential operations on memory qubits}:
            \State $\epsilon_3 \gets W_{ent}^M$
            \State Entangle memory qubits: CRZ($M_1$, $M_2$, $\epsilon_3$)
        \EndFor
        \State Measure and reset qubits in \textit{Readout} register to $\ket{0}$.
    \EndFor
    \State \textbf{Return:} Quantum Circuit
\EndProcedure
\end{algorithmic}
\label{qesnalgo}
\end{algorithm}
Randomly initialized weights gates play an important role in creating reservoir-like behavior in the QC. This approach mimics the inherent randomness and sparsity of ESNs, but with quantum computers. Weights are scaled by the number of qubits, size of the context window, and the number of repeated blocks within the QC to ensure that any single data point does not greatly perturb the system. The distribution for all weights in the QC are centered around a value in $(0, \pi]$ to ensure there is some perturbation happening at each time step. Sparsity is injected into the two qubit gate connections, making most of the entanglement weights zero, mimicking the sparse connectivity found in classical ESNs, and reduces the number of potential gate errors. 

\subsection{Data Handling and Training}\label{subsec:data-training}
Training data is created from numerical simulations of the Lorenz system, a well-known chaotic system often used to test the performance of predictive models. The Lorenz system is defined by the following set of differential equations:
\begin{align*}\label{eq:lorenz}
    \frac{\rm{d}x}{\rm{d}t} = \sigma (y - x), \quad
    \frac{\rm{d}y}{\rm{d}t} = x (\rho - z) - y, \quad
    \frac{\rm{d}z}{\rm{d}t} = xy - \beta z \text{,}
\end{align*}
where $\sigma = 10$, $\beta = \frac{8}{3}$, and $\rho = 28$ are parameters that define the behavior of the system. The training set consists of 9900 data points, split into training and test sets of 6900 and 3000 data points, respectively, where each data point contains the $x(t)$, $y(t)$, and $z(t)$ variables of the Lorenz system. In our experiments, only the $x(t)$ was fed into the QESN, with the task being to predict $y(t)$ and $z(t)$ given this single signal. This setup tests the QESN's ability to learn and predict data generated by complex, nonlinear dynamics. 

First, the QESN circuit is run on the training data, which produces a set of output signals from quantum measurements. This output signal is then used to fit a regression model with a washout length of 300, giving adequate time for the reservoir to reach an equilibrium state. Finally, the QESN circuit is run on the test data, and the learned weights are used to perform predictions. Elastic net regularization was used to prevent overfitting for each test case, with the parameters being tuned separately for each different test case with the goal being to minimize test error rate.

\section{Results and Analysis}\label{sec:results}
Results are gathered from mid-circuit measurements taken and recovered in two different ways: (a) as expectation values that give us just the mean value of each qubit's Pauli-Z outcome, and (b) as the entire probability distributions over the computational basis, which give {$2^{\frac{n_q}{2}}$} features, {where $\frac{n_q}{2}$ is the number of readout qubits}. These mid-circuit measurements are accumulated over 60,000 shots for a single circuit run, ensuring statistical robustness and accuracy of the measured outputs. {Taking a large number of shots of the circuit is extremely important as sampling errors can propagate throughout the network at each time step}. The QESN's performance is evaluated based on its ability to accurately predict the ($y, z$) components of the Lorenz system using $x(t)$, as this task typically requires a fading memory and nonlinear activation function in classical reservoir networks.
   \begin{figure}[h]
        \centering
        \includegraphics[width=\linewidth, keepaspectratio,trim={0cm 0 0 1.5cm},clip]{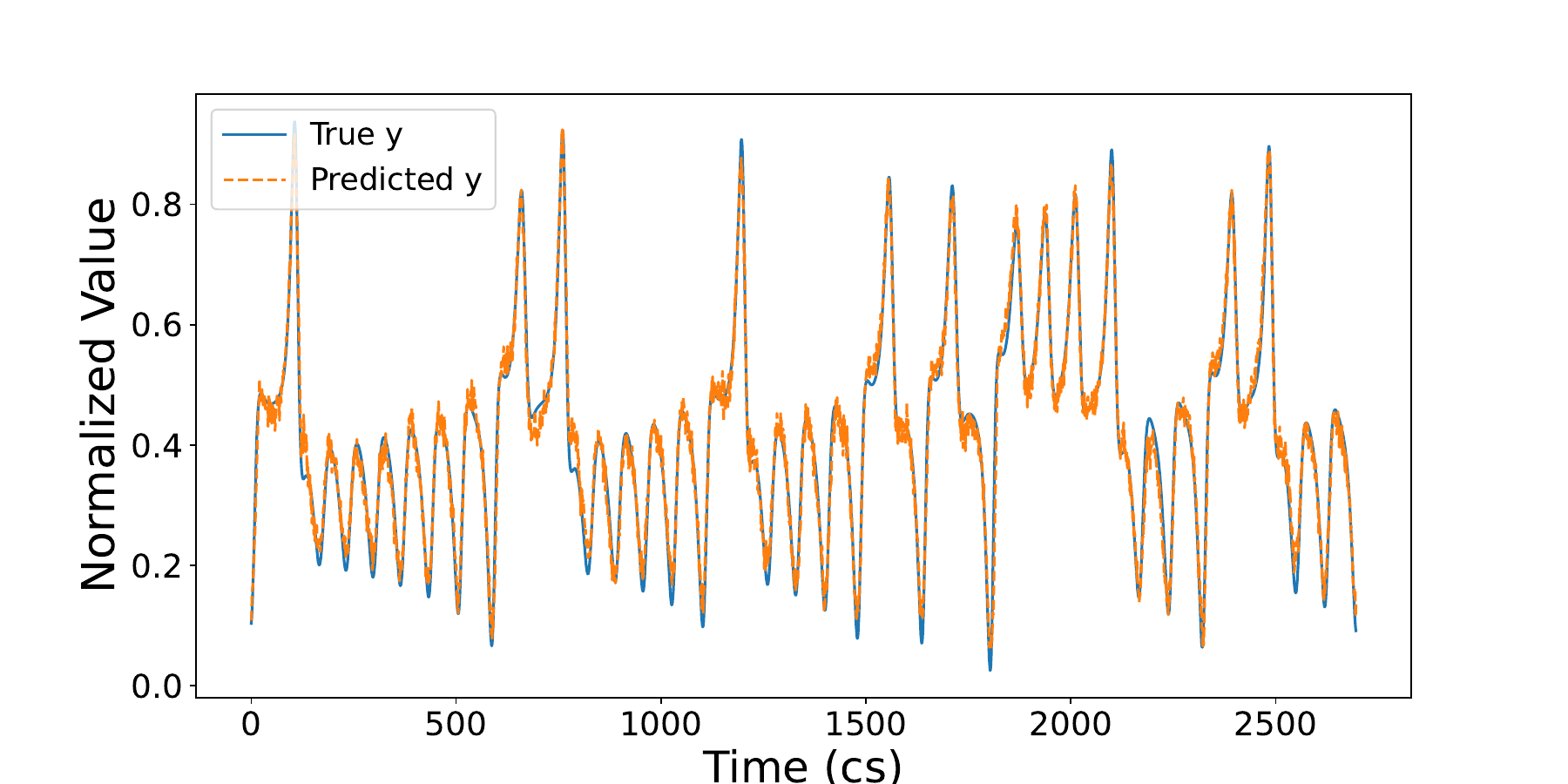}
        \caption{\footnotesize Test prediction from X to Y component in the Aer simulator with 16 qubits.}
        \label{fig:qrnn_test_y}
    \end{figure}
    \begin{table}[h]
        \centering
        \small 
        \begin{tabular}{c|c c c}
             Qubits & Expectation & Distribution & Distribution\\
              & Value &  & w. Noise\\
             \hline
             \textbf
{4 Qubits} &  & \\
             Training RMSE & \textbf{.1124} & .1177 & .1468\\
             Test RMSE & \textbf{.1112} & .1185 & .2016\\
             \hline
             \textbf{6 Qubits} & & \\
             Training RMSE & .0986 & \textbf{.0616} & .1193\\
             Test RMSE & .0963 & \textbf{.064} & .1315\\
             \hline
             \textbf{8 Qubits} &  & \\
             Training RMSE & .0822 & \textbf{.0429} & .1110\\
             Test RMSE & .0798 & \textbf{.0463} & .1285\\
             \hline
             \textbf{10 Qubits} &  & \\
             Training RMSE & .0688 & \textbf{.0425} & .1258\\
             Test RMSE & .0699 & \textbf{.0422} & .1298\\
             \hline
             \textbf{12 Qubits} &  & \\
             Training RMSE & .0631 & \textbf{.0378} & .0986\\
             Test RMSE & .0635 & \textbf{.0377} & .1282\\
             \hline
             \textbf{14 Qubits} &  & \\
             Training RMSE & .0476 & \textbf{.024}  & .0754\\
             Test RMSE & .046 & \textbf{.0249} & .0988\\
             \hline
             \textbf{16 Qubits} &  & \\
             Training RMSE & .0488 & \textbf{.0225} & .0573\\
             Test RMSE & .0493 & \textbf{.0237}  & .0895\\
        \end{tabular}
        \caption{\footnotesize Simulated training and test error using various different feature recovery methods and noise configurations measured in RMSE (Root mean squared error). An \textbf{IBM Fez} noise model was used to gather the noisy results. The best run from each category was used, and the elastic net regularization parameters were tuned for each bin to get lower test loss.}
        \label{tab:comparison_table}
    \end{table}
The noiseless Aer simulator was first used to test our algorithm using differing even number of qubits from 4 to 16. An NVIDIA DGX-A100 was used to collect results. 
The compiled circuit depth for the Aer simulation was about 300k {and the depth for each time step was about 27 depending on the seed for the optimizer and weights}. For statistical robustness, several seeds were tested and the best result from each bin was taken. Later, a noise model was compiled based off of the \textbf{IBM Fez} quantum computer to test real world applicability, with future results aiming to be run on the actual hardware. 

\begin{figure}[h]
        \centering
        \includegraphics[width=\linewidth, keepaspectratio,trim={0cm 0 0 1.4cm},clip]{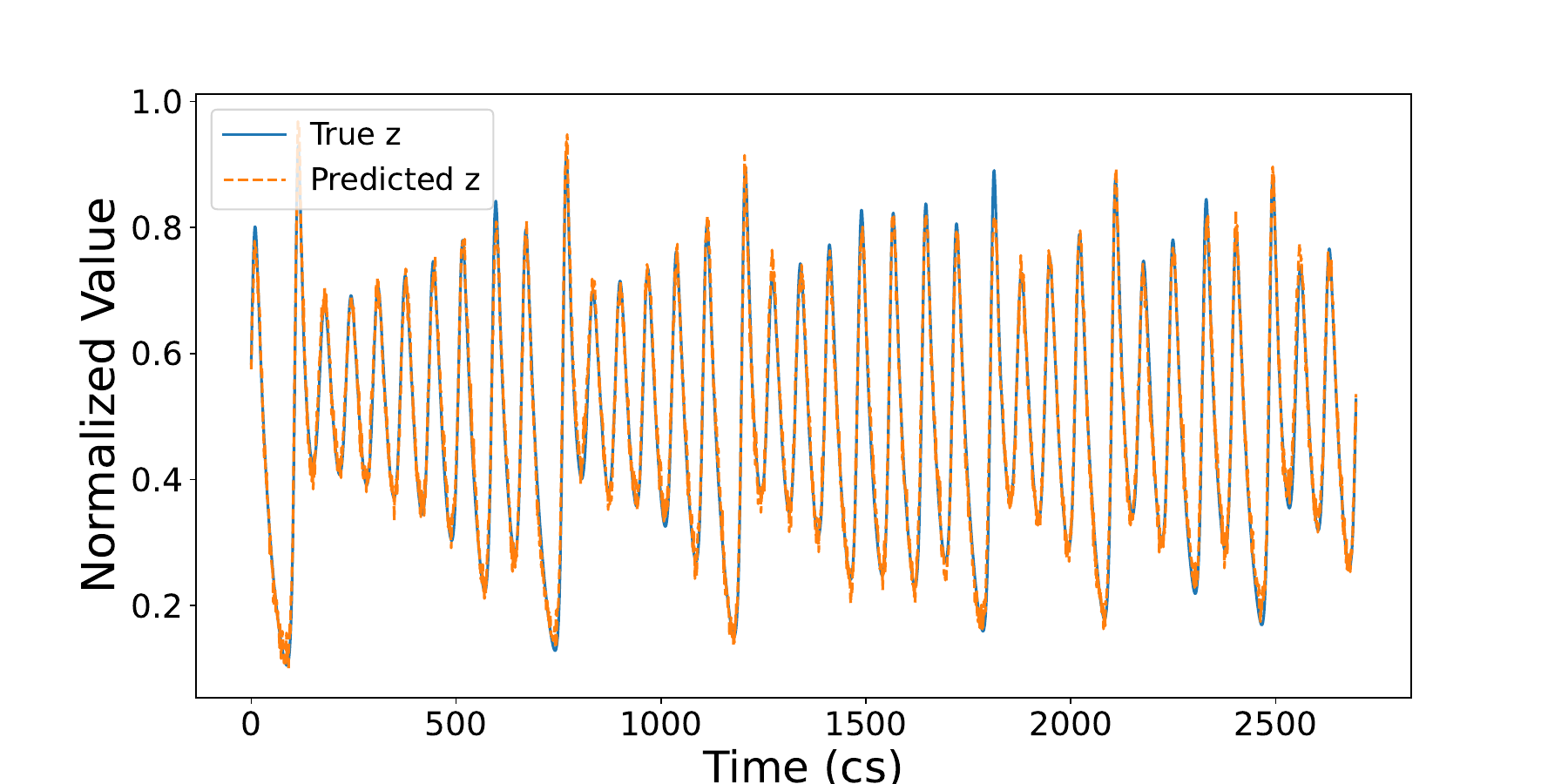}
        \caption{\footnotesize Test prediction from X to Z component in the Aer simulator with 16 qubits.}
        \label{fig:qrnn_test_z}
    \end{figure}

Table \ref{tab:comparison_table} shows a comparison of using expectation values vs the entire probability distribution for various different number of qubits. Predictions of the $y$ and $z$ components were also presented with the predicted signal overlayed on the actual signal in Figures \ref{fig:qrnn_test_y} and \ref{fig:qrnn_test_z}. It is shown that using the entire probability distribution of the output is better in terms of RMSE compared to the expectation value. We also show that the current IBM Fez noise model produces promising results in the ``Distribution w. Noise'' test. It is expected that the noise model will perform worse, but as quantum computers continue to improve, we expect that the difference between noiseless and noisy simulations will be reduced.
\subsection{Comparison with Classical Techniques}\label{subsec: comparison with classical ESN}
\begin{figure}[H]
    \centering
    \includegraphics[width=\linewidth, keepaspectratio,trim={0cm 0 0 1.4cm},clip]{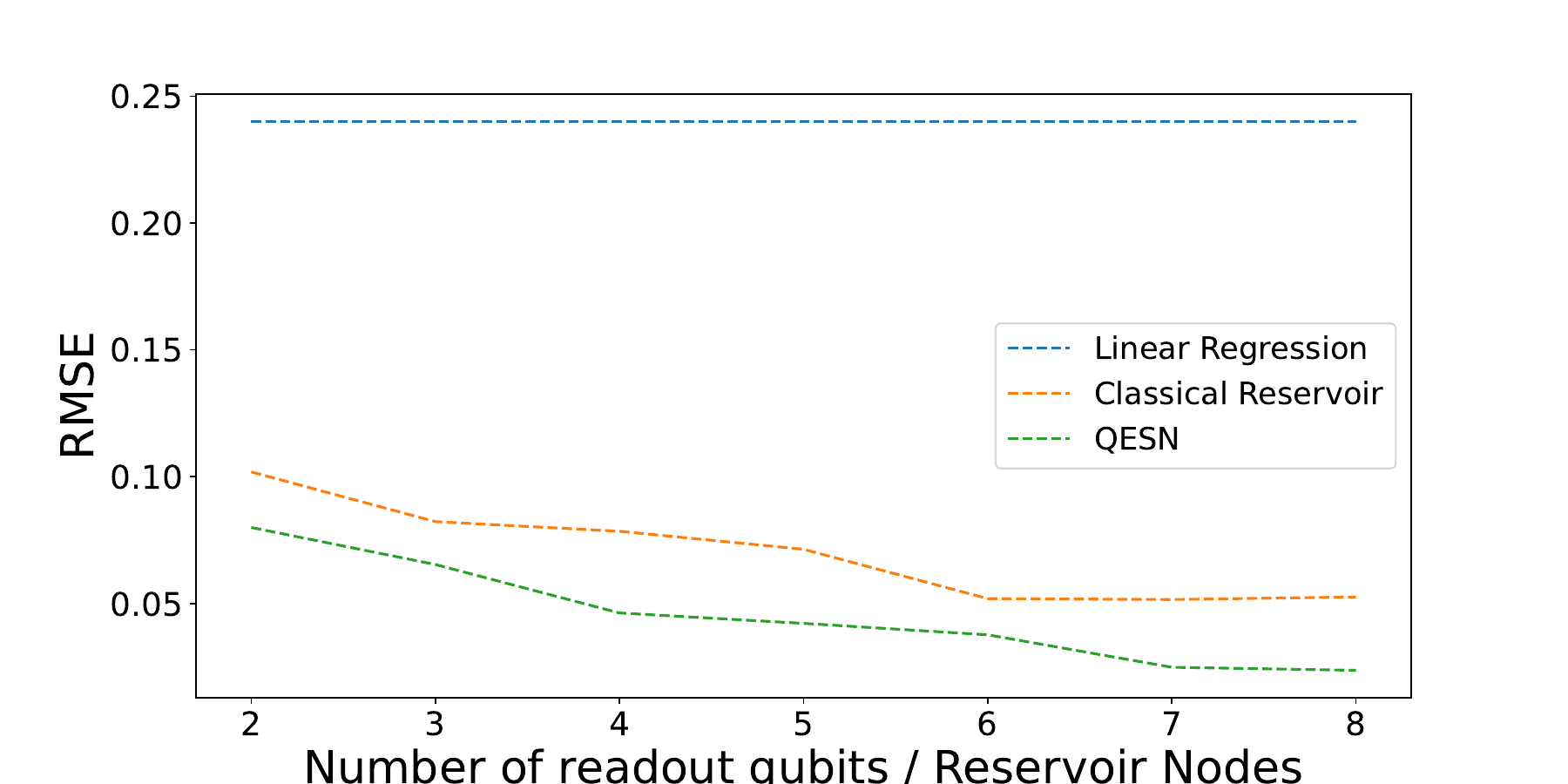}
    \caption{\footnotesize Test set loss with classical reservoir, basic linear regression, and QESN architecture.}
    \label{fig:qrnn_comparison_aer}
    \vspace{-0.3cm}
\end{figure}

A classical ESN was trained and tested in direct comparison to the QESN. 
While there is always room for better hyperparameter tuning, in our results we consistently observed a large performance gain on the noiseless QESN simulation in comparison to the classical ESN when the number of classical ESN reservoir nodes is fixed to the number of readout qubits in the QESN. The test accuracy is shown to be lower on the QESN in every case (Figure \ref{fig:qrnn_comparison_aer}), which could be due to the inherent expressivity of qubits over bits, and the larger memory and feature space that they can provide when entangled.

\section{Conclusions}\label{sec:conclusions}
We have proposed a novel QESN circuit that performs well on the difficult task of time-series prediction in our experiments. This leads us to believe NISQ hardware is almost ready or already ready to perform some general machine learning tasks in the present day. With access to more computational resources and fault tolerant qubits, we expect this architecture could scale beyond classical methods and offer state-of-the-art performance in the solving of PDEs (partial differential equations) and prediction of complex systems. Future research on this method will involve the development of new entanglement schemes, data embedding techniques, and the prediction of more complex systems. 
\section{Acknowledgments}
The authors, EC and VN, would like to acknowledge the funding support from NSWC, Indian Head, under award \# N00174-23-1-0006. Any opinions, findings,
conclusions, or recommendations expressed here are those of the authors and do not necessarily reflect the views of the sponsors.

\newcommand{\BIBdecl}{\setlength{\itemsep}{0em}}
\bibliography{bib}

\end{document}